\documentclass[graybox]{svmult}

\usepackage{mathptmx}       
\usepackage{helvet}         
\usepackage{courier}        
\usepackage{type1cm}        
\usepackage{makeidx}         
\usepackage{graphicx}        
\usepackage{multicol}        
\usepackage[bottom]{footmisc}

\makeindex             

\begin{document}

\title*{Modelling the evolution of galaxies as a function of environment}
\author{Gabriella De Lucia}
\institute{INAF - Astronomical Observatory of Trieste, via G.~B. Tiepolo 11,
  I-34143 Trieste, Italy,
  \email{delucia@oats.inaf.it}
}
%
%
\maketitle

\vskip-1.2truein

\abstract{In this review, I provide an overview of theoretical aspects related
  to the evolution of galaxies as a function of environment. I discuss the main
  physical processes at play, their characteristic time-scales and
  environmental dependency, and comment on their treatment in the framework of
  hierarchical galaxy formation models. I briefly summarize recent results and
  the main open issues.}

\section{A premise on `environment'}
\label{sec:1}

Historically, both theoretical and observational studies trying to assess the
role of environment on galaxy evolution have been focused on galaxy clusters.
One important reason for this is given by the practical advantage of having
many galaxies in relatively small regions of the sky, and all approximately at
the same distance. As laboratories to study galaxy evolution, however, clusters
represent {\it rare} and {\it biased} systems: they originate from the highest
peaks of the primordial density field, and evolutionary processes in these
systems are expected to proceed at a somewhat accelerated pace with respect to
regions of the Universe with `average' density. Only about ten per cent of the
cosmic galaxy population resides in clusters in the local Universe, and this
fraction decreases with increasing redshift. 

The recent completion of large spectroscopic and photometric surveys at
different cosmic epochs has given new impetus to observational studies trying
to assess the role of environment in galaxy evolution. Unfortunately, many
different definitions of {\it environment} are used in the literature
(e.g. some estimate of the halo mass, the number of neighbours counted in some
volume, etc), depending on the available observational data, as well as on their
quality. This unfortunate but inevitable situation makes results from different
surveys and/or at different cosmic epochs difficult to compare, and prevents
them from putting really strong constraints on galaxy formation models.

In addition, it should be noted that in order to establish that physical
processes related to a particular environment do play a role, it would be
necessary to compare the evolution of {\it the same} galaxies in different
environments. And this is certainly a difficult task from the observational
viewpoint. Finally, one should take into account the hierarchical evolution of
cosmic structures: dark matter collapses into haloes in a bottom up fashion,
with small systems forming first and subsequently merging to form more massive
structures. In this framework, galaxies might experience different
`environments' during their lifetimes. So, for example, galaxies residing in a
cluster today might have suffered some degree of `pre-processing' in lower mass
systems like galaxy groups. 

\section{Methods}
\label{sec:2}

Hierarchical galaxy formation models find their seeds in the pioneering work by
\cite{wr78}. Galaxies are believed to originate from the condensation of gas at
the centre of dark matter haloes: during the collapse of cosmic structure, gas
is shock heated to very high temperatures and relaxes to a distribution that
exactly parallels that of dark matter. Gas then cools, primarily via thermal
Bremsstrahlung, and conservation of angular momentum leads to the formation of
a rotationally supported disc. Mergers and instabilities form bulges, that can
eventually grow a new disc, provided the system is fed by an appreciable
cooling flow.

Different techniques are used to link the observed properties of luminous
galaxies to those of the dark matter haloes in which they reside:

\begin{itemize}
\item In {\bf semi-analytic models} of galaxy formation, the evolution of the
  baryonic components of galaxies is followed using simple yet physically
  and/or observationally motivated prescriptions to model complex physical
  processes like star formation, feedback, etc. Modern semi-analytic models
  take advantage of high-resolution N-body simulations to specify the location
  and evolution of dark matter haloes, which are assumed to represent the
  birth-places of luminous galaxies. This method can access large dynamic
  ranges in mass and spatial resolution, and allows a fast exploration of the
  parameter space, and an efficient investigation of different specific physical
  assumptions.

\item {\bf Direct hydrodynamical simulations} provide an explicit description
  of gas dynamics. This method is computationally more expensive then analytic
  models, so that large cosmological simulations are still somewhat limited by
  relatively low mass and spatial resolution. In addition, and perhaps most
  importantly, complex physical processes such as star formation,
  feedback, etc. still need to be modelled as {\it sub-grid physics}, either
  because the resolution of the simulation becomes inadequate or because (and
  this is almost always true) we do not have a `complete theory' for the
  particular physical process under consideration.

\item {\bf Halo Occupation Distribution (HOD) models} are based on a
  statistical characterization of the link between dark matter haloes and
  galaxies, bypassing any explicit modelling of the physical processes driving
  galaxy formation and evolution. 
\end{itemize}

In the following, I will focus primarily on modelling of environmental processes
through the first two methods mentioned above (for more details on HOD models,
I refer to van den Bosch this volume, and references therein). 

\section{Physical processes}
\label{sec:3}

Theoretically, there are a number of physical processes that can affect the
evolution of galaxies in high density environments. In the following, I provide
an overview of these processes, and describe how they are included in
hierarchical galaxy formation models pointing out the main limitations and
problems.

{\bf Galaxy mergers :}
Galaxy mergers and more generally strong galaxy-galaxy
interactions, are commonly viewed as a rarity in massive clusters because of
the large velocity dispersions of the systems.  Mergers are, however, efficient
in the infalling group environment, and may represent an important
`preprocessing' step in the evolution of cluster galaxies.  Numerical
simulations (see e.g. \cite{m04,tt72,bh96,c08} and references therein) have
shown that close interactions can lead to a strong internal dynamical response
driving the formation of spiral arms and, in some cases, of strong bar modes.
Sufficiently close encounters can completely destroy the disc, leaving a
kinematically hot remnant with photometric and structural properties that
resemble those of elliptical galaxies.

\begin{figure}[t]
%
\includegraphics[scale=.37]{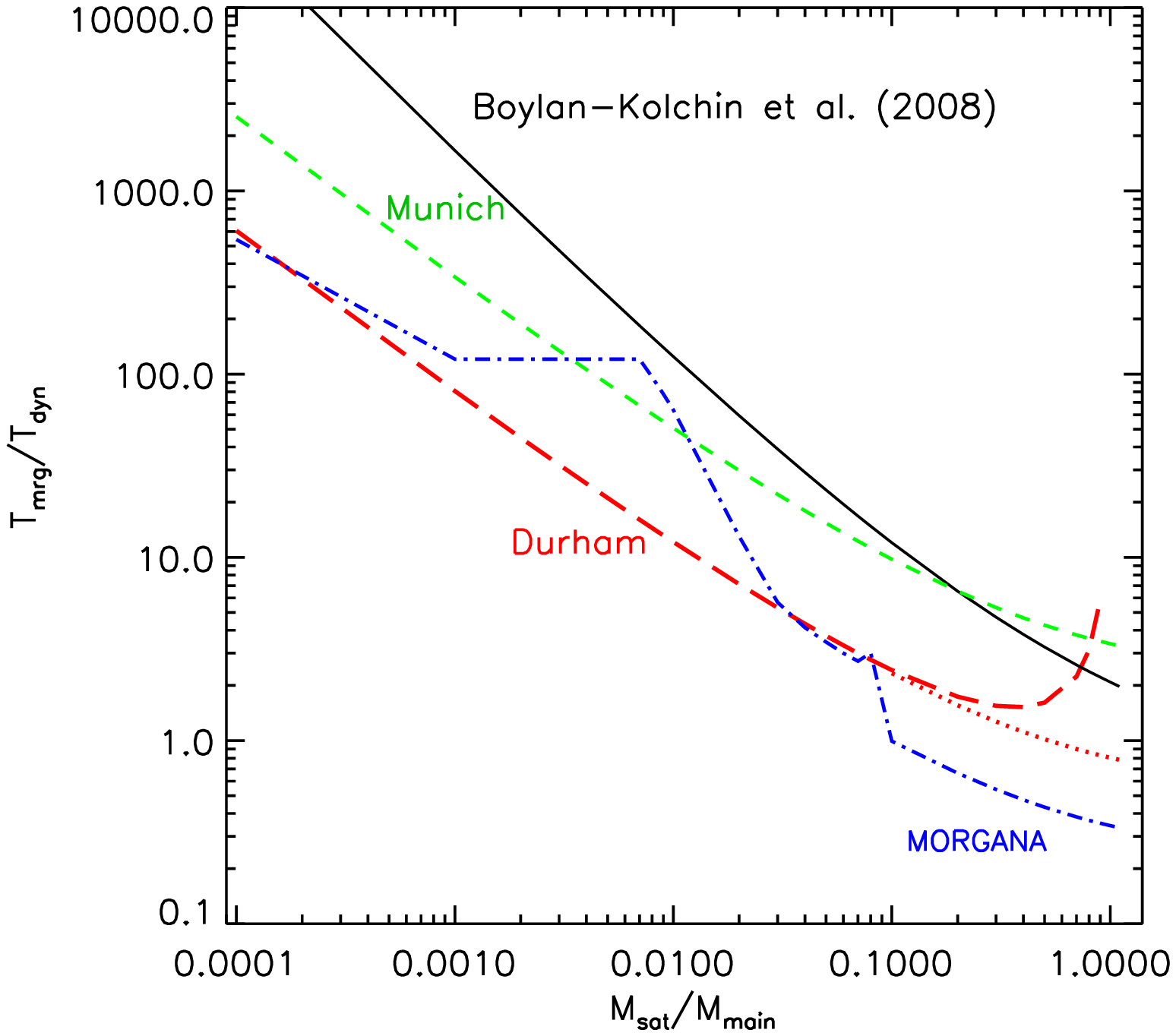}
\includegraphics[scale=.30,angle=-90, bb=554 50 66 575]{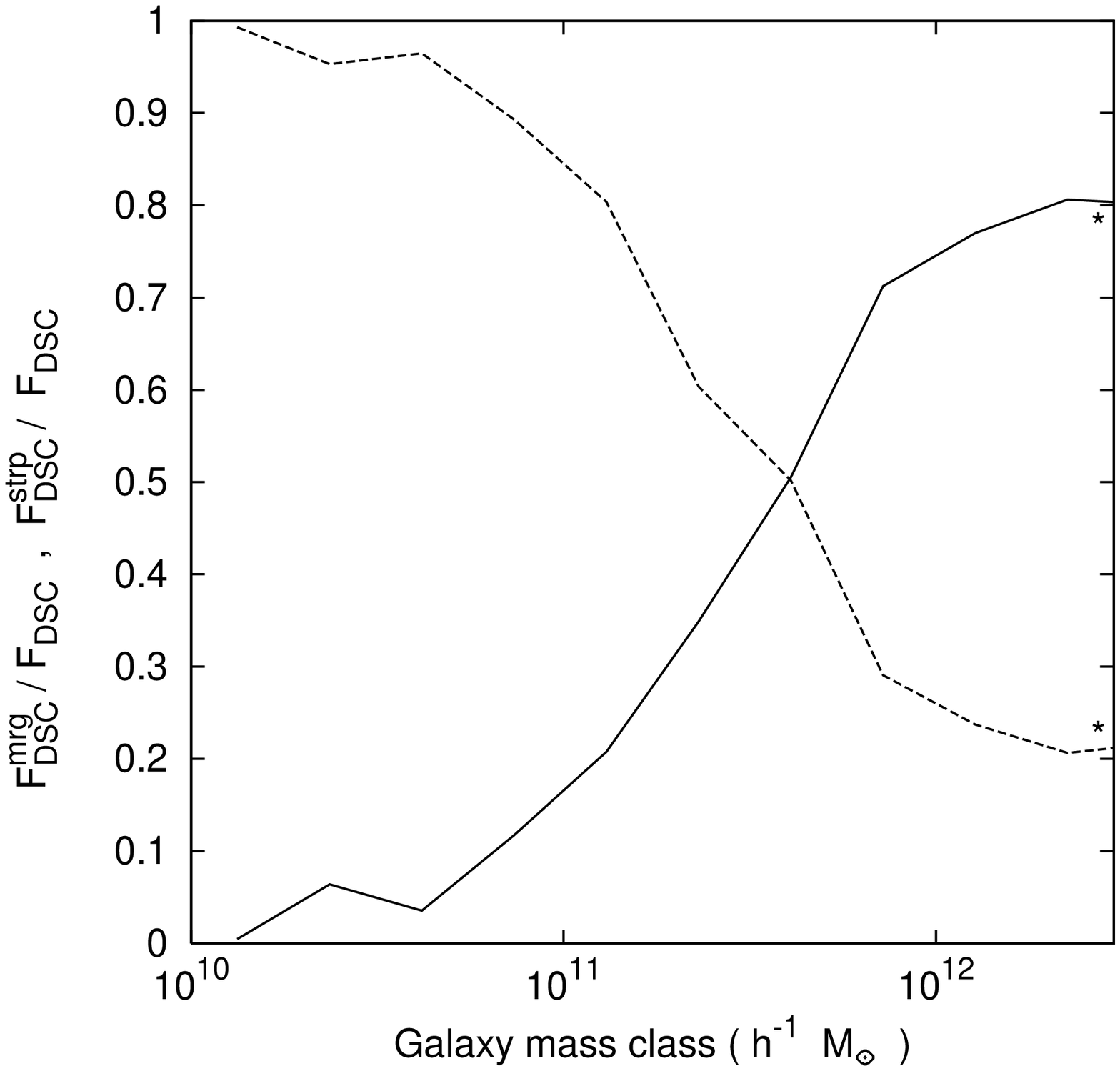}
\caption{{\bf Left:} Merger times (in units of dynamical times) as a function
  of the baryonic mass ratio. Dashed, long-dashed and dot-dashed lines
  correspond to the standard assumptions adopted by the Munich, Durham and
  {\small MORGANA} models respectively. The thick solid line corresponds to the
  fitting formula provided by \cite{bk08}, for circular orbits. {\bf Right:}
  The fraction of diffuse stars arising from mergers associated with the
  assembly history of the BCG (solid line), and with tidal stripping of stars
  from galaxies orbiting in the cluster potential (dashed line).}
\label{fig:mergers}       
\end{figure}

Mergers are intrinsically included in hierarchical galaxy formation models, and
represent an important channel for the formation of bulges (see also Wilman,
this volume). It should be noted that the number of `important' mergers
increases with increasing stellar mass, but is not as large as commonly thought
\cite{dl06}. In semi-analytic models, galaxy mergers are included adopting some
variants of the classical Chandrasekhar dynamical friction formula. The left
panel of Fig.~\ref{fig:mergers} (from \cite{dl10}) compares the merger times
(in units of dynamical times) adopted in three different semi-analytic models
with results from numerical simulations by \cite{bk08}. The figure shows that,
over the range of mass-ratios that provide merger times shorter than the Hubble
time ($M_{\rm sat}/M_{\rm main} > 0.1$), there are large differences between
different models. As discussed in \cite{dl10}, this has important consequences
for the assembly history of massive galaxies, in particular of the brightest
cluster galaxies (BCGs).

Controlled hydrodynamical simulations of merging disk galaxies are used to
`tune' the efficiency of merger driven starbursts in these models
\cite{c08}. One important point, often neglected, is that these simulations are
usually set-up using structural properties of nearby disk galaxies, which might
not necessarily provide a good representation of systems merging at high
redshift.

Recent theoretical work (\cite{m07} - their Fig.~8 is reproduced in the right
panel of Fig.~\ref{fig:mergers}) suggests that merges, and more specifically
those associated with the family tree of the BCG, are responsible for the
largest fraction of the `diffuse stellar component' that is observed in galaxy
clusters. Semi-analytic models have only recently turned their attention to
this component, and often model its formation only through tidal stripping
\cite{s08,g10}.

{\bf Gas stripping :}
Galaxies travelling through a dense intra-cluster medium suffer a strong
ram-pressure stripping that can sweep cold gas out of the stellar disc
\cite{gg72}. Depending on the binding energy of the gas in the galaxy, the
intra-cluster medium will either blow through the galaxy removing some of the
diffuse interstellar medium, or will be forced to flow around the galaxy
\cite{cs77,n82}. Ram-pressure stripping is expected to be more important at the
centre of massive systems because of the large relative velocities and higher
densities of the intra-cluster medium. By considering the distribution and
`history of ram-pressure' experienced by galaxies in clusters, \cite{bdl08}
estimated that strong episodes of ram-pressure are indeed predominant in the
inner core of the clusters. They also showed, however, that virtually all
cluster galaxies suffered weaker episodes of ram-pressure, suggesting that this
physical process might have a significant role in shaping the observed
properties of the entire cluster galaxy population. In addition, \cite{bdl08}
found that ram-pressure fluctuates strongly so that episodes of strong
ram-pressure alternate to episodes of weaker ram pressure, possibly allowing
the gas reservoir to be replenished and intermittent episodes of star formation
to occur. 

While both numerical simulations and analytic studies show that ram-pressure
stripping affects significantly the amount of gas in cluster galaxies, this
physical process is usually not included in semi-analytic models of galaxy
formation, with the exception of a couple of studies \cite{on03,l05}. These
conclude that the inclusion of this physical processes causes only mild
variations in galaxy colours and star formation rates (I will explain why this
is the case below - see `strangulation' section). It should be noted that these
studies include ram-pressure stripping using the original analytic formulation
proposed by \cite{gg72}. Recent numerical work \cite{rb07} has shown that this
formulation often provides incorrect mass-loss rates. In addition, the simple
models used so far do not consider the possibility that ram-pressure stripping
could temporarily enhance star formation \cite{v01,bdl08}.

{\bf Strangulation :} Current theories of galaxy formation assume that, when a
galaxy is accreted onto a larger structure, the gas supply can no longer be
replenished by cooling that is suppressed by the removal of the hot gas halo
associated with the infalling galaxy \cite{l80}. This process is usually
referred to as `strangulation' (or `starvation' or `suffocation').  It is
common to read in discussions related to these physical mechanisms, that
strangulation is expected to affect the star formation of cluster galaxies on
relatively long timescales, and therefore to cause a {\it slow} decline of the
star formation activity.  In semi-analytic models, however, this process is
usually associated to a strong supernovae feedback, and is assumed to be
instantaneous. As a consequence, galaxies that fall onto a larger system
consume their cold gas rapidly, moving onto the red-sequence on very short
time-scales (see e.g. \cite{w07}). This is also why ram-pressure stripping is
not found to have a significant influence in these models: galaxies turn red
and dead so quickly that ram-pressure does not have enough time to affect their
colours or star formation rates.

\begin{figure}[t]
%
\includegraphics[scale=.40]{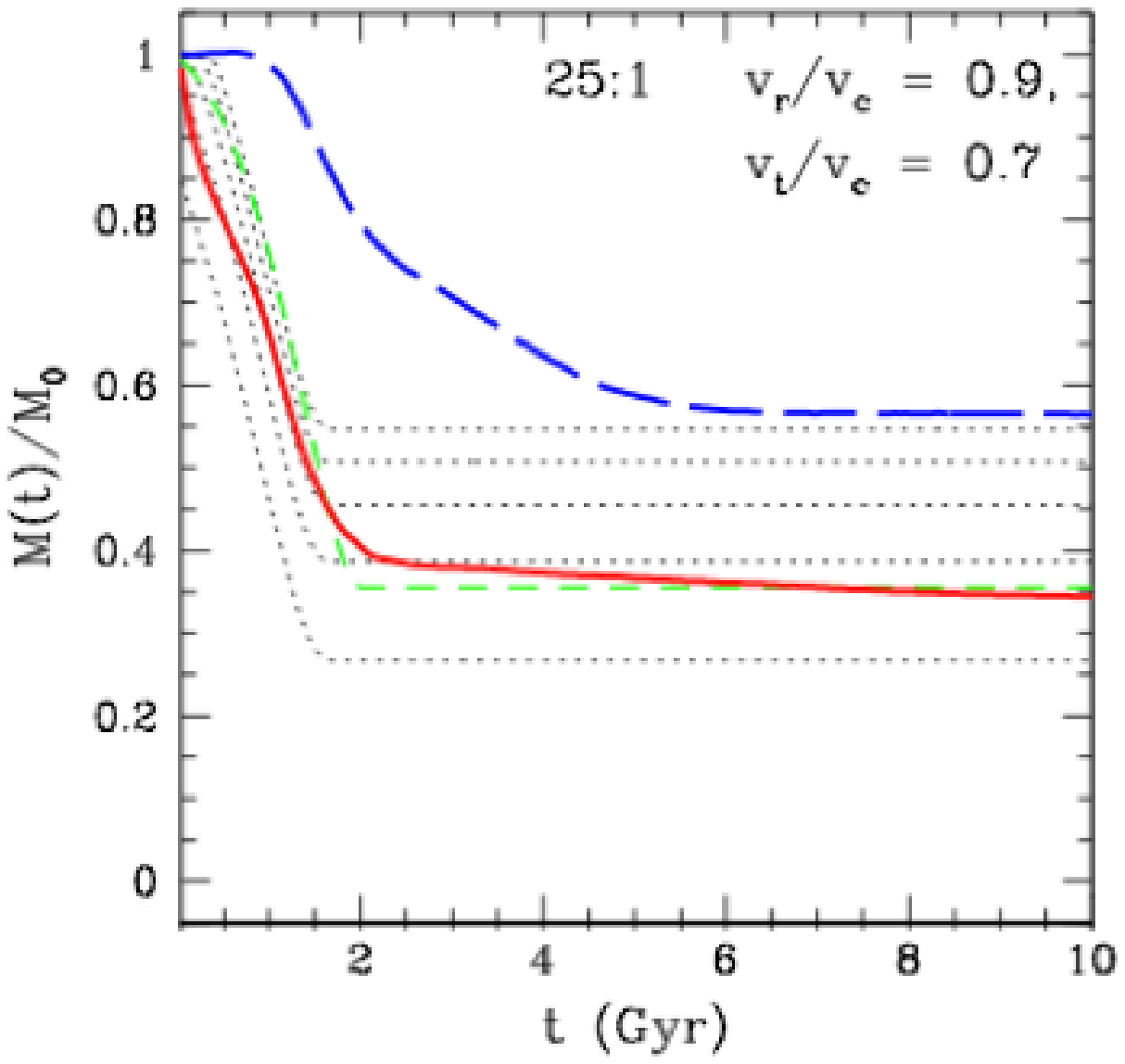}
\includegraphics[scale=.45]{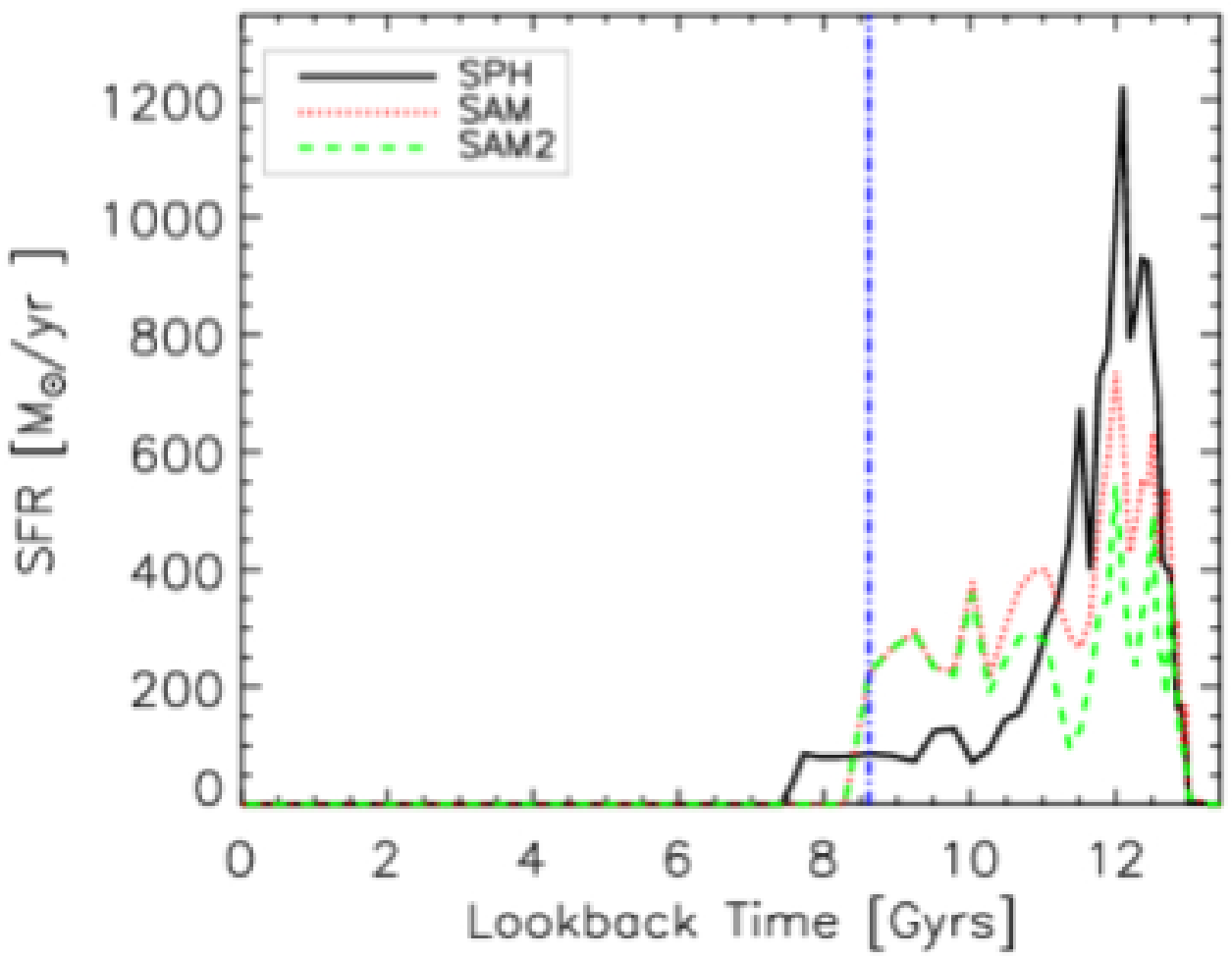}
\caption{{\bf Left:} From \cite{mc07}. The solid and dashed lines show the
  evolution of the bound mass of gas and dark matter in a simulation (see
  original paper for details) of a galaxy accreted onto a larger system. {\bf
    Right:} From \cite{s10}. The star formation history of a galaxy residing in
  a cluster today. The vertical line indicates the time of accretion. The solid
  line is from a hydrodynamical simulations while the dashed and dotted lines
  are from a semi-analytic model (see original paper for details).}
\label{fig:strangulation}       
\end{figure}

Numerical simulations have recently pointed out that the stripping of the hot
halo associated with infalling galaxies should not happen instantaneously. The
left panel from Fig.~\ref{fig:strangulation} is reproduced from \cite{mc07} and
shows the evolution of the dark matter (solid line) and gas component (dashed
line) associated with a galaxy placed on a realistic orbit through a cluster
(see original paper for details). This particular simulation shows that the
galaxy can retain $\sim 30$ per cent of its initial hot gas even $\sim 10$~Gyr
after accretion. The right panel is reproduced from \cite{s08} and compares the
star formation history of a cluster satellite galaxy from hydrodynamical
simulations and semi-analytic models. \cite{s08} find that some cooling occurs
on satellites and that this can last up to $\sim 1$~Gyr after accretion, but
this seems to be important only for the most massive satellites. These
simulations provide important (albeit still inconclusive) inputs for a revised
treatment of this physical process in the framework of semi-analytic models of
galaxy formation (see e.g \cite{f08,w10}).

{\bf Harassment :} Galaxy harassment is a process that is not usually included
in semi-analytic models of galaxy formation. The process has been discussed in
early work on the dynamical evolution of cluster galaxies \cite{r76}, and has
been explored in detail using numerical simulations \cite{fs81,m98}. These have
shown that repeated fast encounters, coupled with the effects of the global
tidal field of the cluster, can drive a strong response in cluster
galaxies. The efficiency of the process is, however, largely limited to
low-luminosity hosts, due to their slowly rising rotation curves and their
low-density cores. Therefore, it is believed that harassment might have an
important role in the formation of dwarf ellipticals or in the destruction of
low-surface brightness galaxies in clusters \cite{m05}, but it is less able to
explain the evolution of luminous cluster galaxies.

\begin{figure}[t]
%
\includegraphics[scale=.40]{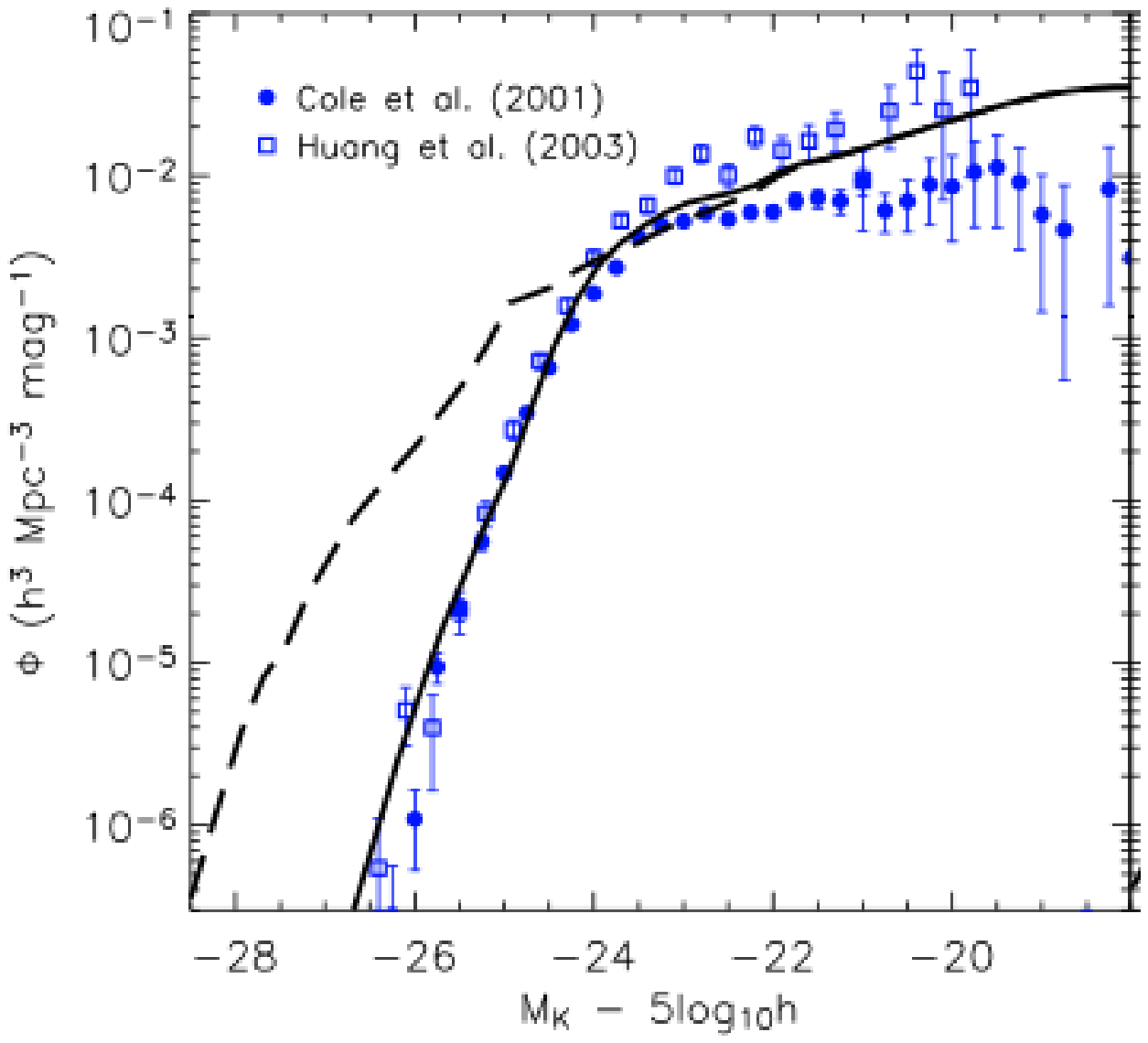}
\includegraphics[scale=.45]{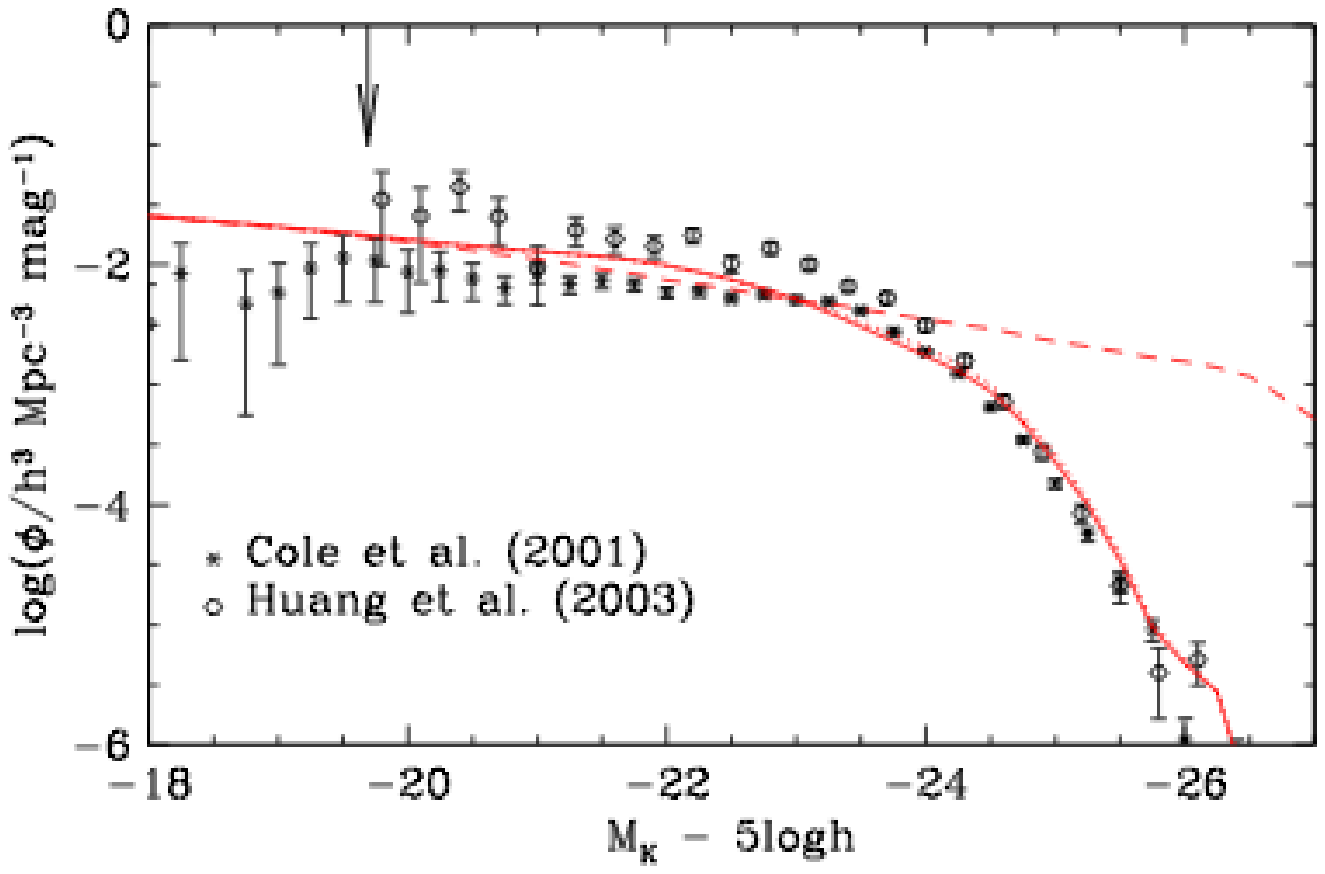}
\caption{{\bf Left:} The predicted K-band luminosity function from two recently
  published semi-analytic models (\cite{c06} left panel and \cite{b06} right
  panel). In both panels, the solid and dashed lines show model predictions
  with and without AGN feedback, respectively.}
\label{fig:lf}       
\end{figure}

{\bf AGN heating: } Since the milestone paper by \cite{wf91}, it has been
realized that some physical process is needed to suppress cooling flows at the
centre of relatively massive haloes. Early semi-analytic models introduced
ad-hoc prescriptions to suppress cooling flows in haloes above a critical mass.
Modern models have included more accurate and physically motivated
prescriptions, and have confirmed that AGN heating is indeed important to
reproduce the exponential cut-off at the bright end of the galaxy luminosity
function, and the old stellar populations observed for massive galaxies
\cite{dl06}. Fig.~\ref{fig:lf} shows, for example, the predicted K-band
luminosity function from two recently published semi-analytic models
\cite{b06,c06} with and without AGN feedback (solid and dashed lines,
respectively). The prescriptions adopted in these models are, however,
necessarily very schematic and not well grounded in observation. A recent work
by \cite{f10}, in particular, has pointed out that the distributions of radio
sources predicted by recent semi-analytic models is in disagreement with
observational data, confirming that much work still needs to be done in order
to understand exactly how and when AGN feedback is important.

\section{What next?}
\label{sec:5}

The above discussion highlights that there are several areas where our galaxy
formation models could and should be improved. In particular:

\begin{itemize}
\item Different treatments of galaxy mergers lead to merger times that can
  differ up to one order of magnitude \cite{dl10}. This has important
  consequences for the assembly history of the most massive galaxies, and for
  the evolution of the bright/massive end of the luminosity/mass function.
\item Only recently, galaxy formation models have turned their attention to the
  formation of the diffuse stellar component. The increasing amount and quality
  of observational data on the intra-cluster light, and detailed comparisons
  with theoretical models, will provide important constraints on how this
  component forms and evolves as a function of cosmic time.
\item Much recent work has focused on improving our treatment of gas stripping
  from satellite galaxies \cite{g10,w10}. These models, however, are not
  without problems: e.g. the model discussed in \cite{g10} still predicts a
  larger passive fraction among low-mass galaxies than is observed, and an
  excess of intermediate to low mass galaxies beyond $z \sim 0.5$. There
  results call for a deep revision of our modelling of the evolution of
  satellite galaxies.
\item `Radio-mode' AGN feedback represents an elegant solution to a number of
  long-standing problems related to the evolution of massive galaxies. Yet, our
  modelling of this physical process is still very schematic. In particular, the
  strong dependency on halo mass usually adopted in semi-analytic models might
  not be supported by observational data \cite{f10}.
\end{itemize}

To conclude, I would like to remind that the importance of the physical
processes discussed above has been investigated in detail using numerical
simulations, at the typical velocity dispersions of galaxy clusters. Detailed
studies at smaller scales (those of galaxy groups) are often lacking, even
though groups likely represent the most common environment galaxies experience
during their lifetimes.  

The increasing amount and quality of the data being collected in these years
are going to continuously challenge available and future models. This close link
between theory and observations needs to be maintained (and strengthened) in
order to improve our understanding of the processes driving galaxy formation
and evolution as a function of the environment.

\begin{acknowledgement}
  I wish to thank A. Pasquali and I. Ferreras for the organization of a very
  lively and stimulating meeting. I acknowledge financial support from the
  European Research Council under the European Community's Seventh Framework
  Programme (FP7/2007-2013)/ERC grant agreement n. 202781.
\end{acknowledgement}
%


\end{document}